\documentclass{optica-article}

\journal{opticajournal} % for journals or Optica Open

\articletype{Research Article}

\usepackage{lineno}
\usepackage{amsmath}
\newcommand{\U}{\ensuremath{\mathrm{U}}}
\newcommand{\V}{\ensuremath{\mathrm{V}}}

\newcommand{\iN}{\ensuremath{\mathcal{N}}}
\newcommand{\mS}{\ensuremath{\mathbf{S}}}
\newcommand{\mI}{\ensuremath{\mathbf{I}}}

\begin{document}

\title{Single-Frequency Symmetry-Empowered Through-Barrier Sensing in Reconfigurable Complex Media}

\author{Shuai S. A. Yuan\authormark{1}, Viktar Asadchy\authormark{1}, and Philipp del Hougne\authormark{1,2,*}}

\address{\authormark{1}Department of Electronics and Nanoengineering, Aalto University, 00076 Espoo, Finland\\
\authormark{2}Univ Rennes, CNRS, IETR - UMR 6164, F-35000, Rennes, France}

\email{\authormark{*}philipp.del-hougne@univ-rennes.fr} %% email address is required; see note below about the corresponding author designation

% use {asbstract*} to suppress the copyright line. Copyright information will be added in production

\begin{abstract*} 
Mirror symmetry can strongly enhance the transmission of waves through a barrier inside a complex medium. We recently showed that this phenomenon enables quantitative through-barrier sensing: by tuning programmable scatterers on one side of the barrier to maximize the \textit{broadband} total transmission through the barrier, the characteristics of scatterers at mirror-symmetric positions on the other side of the barrier can be determined. Considering a sufficiently large bandwidth was crucial to ensure that no accidental narrowband asymmetric resonance can outperform the symmetry-induced transmission enhancement. Here, we overcome this scheme's vexing need for a large bandwidth by replacing the underlying frequency diversity with configurational diversity. Specifically, we introduce auxiliary tunable scatterers at mirror-symmetric positions on either side of the barrier and sweep their characteristics through a series of random mirror-symmetric configurations. We tune the programmable main scatterers on one side of the barrier to maximize the average of the total through-barrier transmission over a series of configurations of the auxiliary scatterers \textit{at a single frequency}, in order to sense the characteristics of the main scatterers on the other side of the barrier. We systematically study the accuracy of our single-frequency sensing scheme based on a multiport-network system model that cascades two mirror-related wave-chaotic cavities with a weakly transmitting barrier in between. We further examine an extension to non-reciprocal chaotic cavities involving circulators. Altogether, our results establish configurational diversity as a route to single-frequency, symmetry-empowered through-barrier sensing in reconfigurable complex media.
\end{abstract*}

%%%%%%%%%%%%%%%%%%%%%%%%%%  body  %%%%%%%%%%%%%%%%%%%%%%%%%%
\section{Introduction}

Symmetry can strongly influence wave propagation in complex media. A striking phenomenon occurs in barrier-separated systems with left-right mirror symmetry. When an opaque barrier is placed between two complex media that are mirror images of each other, transmission through the barrier can be strongly enhanced compared to the asymmetric case. This effect has been demonstrated in symmetric chaotic double quantum dots, where mirror symmetry boosts conductance across a tunnel barrier~\cite{macucci2007tunneling,whitney2009huge,totaro2010effect,macucci2020optimization}, in graphene-based double dots~\cite{marconcini2013symmetry}, and in symmetric diffusive disordered slabs separated by opaque barriers~\cite{cheron2019broadband,cheron2020broadband,cheron2020sensitivity,davy2021experimental}. The enhancement is a broadband effect arising from constructive interference between mirror-related multiple-scattering paths~\cite{whitney2009huge,cheron2019broadband,borcea2024enhanced,flegontov2025symmetry}.

The symmetry-induced through-barrier transmission enhancement is sensitive to symmetry defects~\cite{whitney2009huge,totaro2010effect,macucci2020optimization,cheron2020broadband,cheron2020sensitivity}. This sensitivity can be leveraged as a sensing mechanism. Early works already pointed in this direction for applications in defect detection~\cite{whitney2009huge,cheron2019broadband} and quantitative single-parameter estimation~\cite{macucci2007tunneling,macucci2020optimization}. Recently, we 
demonstrated an application to quantitative through-barrier sensing of a multi-parameter hidden configuration~\cite{yuan2026symmetry}. The basic principle is to optimize a programmable region on the accessible side of the barrier to restore the mirror symmetry with the unknown region behind the barrier by maximizing the total through-barrier transmission. Crucially, to avoid that the symmetry-induced enhancement could be outperformed by an asymmetric resonant enhancement, this technique requires maximizing the total through-barrier transmission across a sufficiently broad bandwidth~\cite{yuan2026symmetry}. 

In this paper, we demonstrate a single-frequency variant that overcomes the vexing broadband requirement of the scheme proposed in Ref.~\cite{yuan2026symmetry}. At the core of our approach is a replacement of the frequency diversity used in Ref.~\cite{yuan2026symmetry} by configurational diversity, taking inspiration from analogous developments in the field of computational meta-imaging. To overcome the need for a large number of expensive and power-hungry radio-frequency chains, computational meta-imagers multiplex wireless signals across a diverse set of measurement modes onto a significantly reduced number of receivers. Initial proposals relied on frequency multiplexing, leveraging the frequency-selective dispersion of metamaterial apertures~\cite{hunt2013metamaterial,fromenteze2015computational,gollub2017large}. By scanning across a sufficiently broad bandwidth, enough independent frequency points were accessible. Later, programmable metasurfaces enabled the multiplexing of wireless signals across different metasurface configurations at the same frequency~\cite{sleasman2015dynamic,li2016transmission}, thus replacing frequency diversity with configurational diversity. 
A similar development also occurred in the recently emerging field of wireless multiport sensing: while the initial demonstration of retrieving over-the-air the full scattering matrix of a device under test required multiple-input multiple-output measurements with multiple accessible antennas and RF chains~\cite{del2025wireless}, configurational diversity later enabled a single-input single-output implementation in which multiple programmable load states are used~\cite{del2026low}.

We investigate our idea for single-frequency, symmetry-empowered through-barrier sensing leveraging configurational diversity based on a system that deliberately remains close to the one used in Ref.~\cite{yuan2026symmetry}. Our setup here also consists of two mirror-symmetric wave-chaotic cavities containing programmable point scatterers, with an unknown configuration on one side and a controllable configuration on the other. The key difference is a second set of tunable auxiliary scatterers placed at mirror-symmetric positions on both sides of the barrier and always switched in mirror-symmetric configurations; these auxiliary tunable scatterers generate the configurational diversity that replaces the frequency diversity used in Ref.~\cite{yuan2026symmetry}.
A second difference with respect to Ref.~\cite{yuan2026symmetry} concerns the barrier itself. Rather than coupling the two cavities through a narrow slit in a common wall, we represent the barrier as a multiport scattering network that connects mirror-related cavity ports through weakly transmitting propagating channels. This formulation makes the full system a cascade of three multiport systems: the left cavity, the barrier network, and the right cavity. Besides making the modeling computationally more convenient, this implementation clarifies that the symmetry-induced through-barrier enhancement is not tied to evanescent coupling through the barrier. Similar to Ref.~\cite{yuan2026symmetry}, our present work can be understood as realizing a discrete-basis form of symmetry-empowered through-barrier imaging.

Our contributions are summarized as follows.
\textit{First}, we introduce symmetry-preserving configurational averaging as a replacement for the broadband frequency averaging of Ref.~\cite{yuan2026symmetry}, thereby enabling single-frequency through-barrier sensing.
\textit{Second}, using a single full-wave simulation of a 100-port chaotic cavity together with a computationally efficient cascaded system model based on multiport-network theory (MNT), we systematically quantify how the sensing performance depends on the number of auxiliary tunable scatterers and the number of considered configurations of the auxiliary tunable scatterers. 
\textit{Third}, we extend the framework to non-reciprocal cavities involving ideal circulators and show that mirror-transformed circulator bias preserves the generalized mirror symmetry, whereas non-mirror-transformed bias acts as a symmetry-breaking perturbation that is largely averaged out in the sensing decision.

\section{System Model}
\label{sec:system_model}

In this section, we formulate our single-frequency system model. As in Ref.~\cite{yuan2026symmetry}, the static components of the setup are perfectly mirror-symmetric such that symmetry breaking can only originate from the configuration of the tunable components. Thereby, our system is amenable to computationally efficient multiport modeling, while the underlying symmetry-based principle is, of course, independent of how the asymmetry is physically realized. 
As mentioned, compared to Ref.~\cite{yuan2026symmetry}, the two main differences are that we introduce additional tunable loads to access configurational diversity, and that we treat the system as the cascade of three multiport subsystems. We first describe the considered setup. Then, we
determine the scattering matrix for each of the three involved subsystems in turn. Finally, we determine the system's transmission matrix by evaluating the cascade of the three subsystems.

\subsection{Setup Overview}
\label{subsec_SetupOverview}

\begin{figure*}
    \centering
    \includegraphics[width=0.9\textwidth]{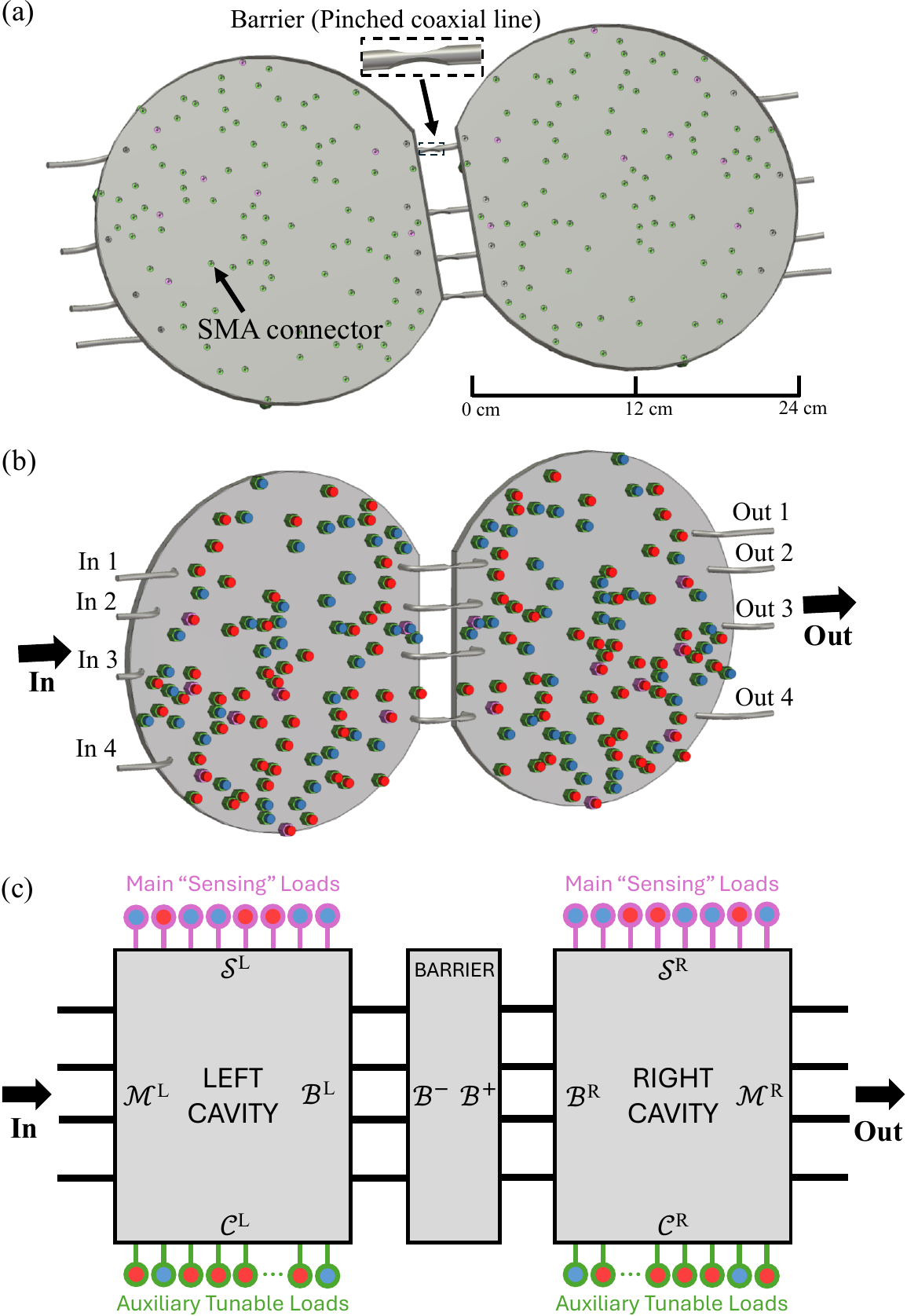}
	\caption{(a,b) Considered setup viewed from above with the cover removed to show the interior in (a), and from below in (b), where the tunable loads -- open circuits (blue) and short circuits (red) -- as well as coaxial cables are visible. The barrier is implemented as weakly transmitting coaxial links between mirror-related port pairs; physically, a weakly transmitting coaxial link can be understood as a ``pinched'' coaxial cable with a local constriction, as illustrated in the inset.
    (c) MNT system model obtained by cascading the left cavity, the barrier, and the right cavity. The port index sets used in the main text are indicated. }
    \label{Fig1}
\end{figure*}

The considered setup is sketched in Fig.~\ref{Fig1}(a,b); geometric details are described in Sec.~\ref{sec_SetupProcedure}. 
Our setup consists of two identical, reciprocal, quasi-two-dimensional D-shaped copper cavities that are related by mirror symmetry and connected by a set of weakly transmitting idealized coaxial links between mirror-related ports. 
D-shaped cavities are widely used examples of complex media because their ray dynamics gives rise to wave chaos~\cite{draeger1997one,ree1999classical,doya2001optimized,redding2015low,faul2025agile}. 
The cavity height is chosen such that only the fundamental vertical mode is supported at the operating frequency. 
Each cavity has multiple ports, all of which are realized by deeply subwavelength SubMiniature version A (SMA) connectors whose center pins protrude into the cavities; in microwave-network terminology, these SMA connectors are electrically small lumped ports. There are four distinct sets of ports, each with a different role.
\textit{First}, $M$ ports are used to inject waves into the left cavity, and the mirror-related $M$ ports of the right cavity are used to receive waves. 
\textit{Second}, $N_\mathrm{B}$ ports of the left cavity are connected pairwise by the weakly transmitting coaxial links to the mirror-related $N_\mathrm{B}$ ports of the right cavity. 
\textit{Third}, $N_\mathrm{S}$ ports in each cavity are terminated by individual loads. 
The load states of these $N_\mathrm{S}$ ``sensing'' ports in the right cavity are static and unknown, whereas the load states of the mirror-related $N_\mathrm{S}$ ``sensing'' ports in the left cavity are known and tunable; the objective of the sensing task is to infer the former by optimizing the latter. 
\textit{Fourth}, another set of $N_\mathrm{C}$ ports in each cavity is terminated by individual loads. These loads are switched in mirror-symmetric pairs on the two sides of the barrier and provide the configurational diversity that replaces the frequency diversity used in Ref.~\cite{yuan2026symmetry}. 
The considered port terminations are open-circuit and short-circuit loads, corresponding to reflection coefficients $+1$ and $-1$, respectively. 
Altogether, the only possible source of mirror-symmetry breaking is therefore a mismatch between the load states of mirror-related ``sensing'' ports.

\subsection{Left Cavity}

The left cavity has $N=M+N_\mathrm{B}+N_\mathrm{S}+N_\mathrm{C}$ ports and is hence characterized by an $N\times N$ scattering matrix $\mathbf{S}^\mathrm{L}$. The sets $\mathcal{M}^\mathrm{L}$, $\mathcal{B}^\mathrm{L}$, $\mathcal{S}^\mathrm{L}$, and $\mathcal{C}^\mathrm{L}$ denote the sets of port indices associated with the four types of ports. In the following, we denote by $\mathbf{S}^\mathrm{L}_{\mathcal{X}^\mathrm{L}\mathcal{Y}^\mathrm{L}}$ the block of $\mathbf{S}^\mathrm{L}$ selected by the sets of indices $\mathcal{X}^\mathrm{L}$ and $\mathcal{Y}^\mathrm{L}$, with $\mathcal{X}^\mathrm{L},\mathcal{Y}^\mathrm{L}\in\{\mathcal{M}^\mathrm{L},\mathcal{B}^\mathrm{L},\mathcal{S}^\mathrm{L},\mathcal{C}^\mathrm{L} \} $. 
The ports in $\mathcal{S}^\mathrm{L}$ are terminated by loads whose reflection coefficients are summarized in the $N_\mathrm{S}$-element vector $\mathbf{r}^\mathrm{L}_\mathrm{S}$. The ports in $\mathcal{C}^\mathrm{L}$ are terminated by loads whose reflection coefficients are summarized in the $N_\mathrm{C}$-element vector $\mathbf{r}_\mathrm{C}$. With $\mathcal{O}^\mathrm{L}=\mathcal{M}^\mathrm{L}\cup\mathcal{B}^\mathrm{L}$ and $\mathcal{T}^\mathrm{L}=\mathcal{S}^\mathrm{L}\cup\mathcal{C}^\mathrm{L}$, where $\cup$ denotes the union of index sets, and $\mathbf{\Phi}^\mathrm{L}=\mathrm{diag}([\mathbf{r}^\mathrm{L}_\mathrm{S}, \ \mathbf{r}_\mathrm{C}])$, standard MNT yields the reduced $(M+N_\mathrm{B})$-port scattering matrix $\widetilde{\mathbf{S}}^{\mathrm{L}}$ at the unterminated ports of the left cavity as a function of the loads on the terminated ports:
\begin{equation}
\widetilde{\mathbf{S}}^{\mathrm{L}}
=
\mathbf{S}^{\mathrm{L}}_{\mathcal{O}^{\mathrm{L}}\mathcal{O}^{\mathrm{L}}}
+
\mathbf{S}^{\mathrm{L}}_{\mathcal{O}^{\mathrm{L}}\mathcal{T}^{\mathrm{L}}}
\mathbf{\Phi}^{\mathrm{L}}
\left(
\mathbf{I}_{N_\mathrm{T}}
-
\mathbf{S}^{\mathrm{L}}_{\mathcal{T}^{\mathrm{L}}\mathcal{T}^{\mathrm{L}}}
\mathbf{\Phi}^{\mathrm{L}}
\right)^{-1}
\mathbf{S}^{\mathrm{L}}_{\mathcal{T}^{\mathrm{L}}\mathcal{O}^{\mathrm{L}}},
\label{eq:left_reduced}
\end{equation}
where $N_\mathrm{T}=N_\mathrm{S}+N_\mathrm{C}$ and $\mathbf{I}_{N_\mathrm{T}}$ denotes the $N_\mathrm{T}\times N_\mathrm{T}$ identity matrix. A detailed derivation of \eqref{eq:left_reduced} can be found in Appendix~B.3 in Ref.~\cite{prod2024updatable}.

\subsection{Barrier}

The barrier is represented as a reciprocal $2N_\mathrm{B}$-port scattering network. 
Its port sets $\mathcal{B}^{-}$ and $\mathcal{B}^{+}$ are connected to the barrier-facing ports of the left and right cavities, respectively. 
We choose the barrier to couple only mirror-related port pairs, with identical transmission amplitude $t_\mathrm{B}$ and reflection amplitude $r_\mathrm{B}$ for each pair:
\begin{equation}
\mathbf{S}^{\mathrm{B}}
=
\begin{pmatrix}
r_\mathrm{B}\mathbf{I}_{N_\mathrm{B}} & t_\mathrm{B}\mathbf{I}_{N_\mathrm{B}}\\
t_\mathrm{B}\mathbf{I}_{N_\mathrm{B}} & r_\mathrm{B}\mathbf{I}_{N_\mathrm{B}}
\end{pmatrix}.
\label{eq:barrier}
\end{equation}
We consider a lossless barrier. Specifically, we choose $t_\mathrm{B}$ to be real and set $r_\mathrm{B}=i\sqrt{1-|t_\mathrm{B}|^2}$, such that $|r_\mathrm{B}|^2+|t_\mathrm{B}|^2=1$ and $\mathbf{S}^{\mathrm{B}}$ is unitary.
Thus, $|t_\mathrm{B}|^2$ directly controls the barrier transparency. 
This scattering-matrix representation implies that the barrier is modeled through propagating channels, excluding evanescent coupling between the cavities.
Physically, one may picture each barrier channel as a coaxial link whose transmission is strongly reduced by a local constriction, i.e., as a ``pinched'' coaxial connection between a pair of mirror-related cavity ports.
Our barrier model can be understood as the discrete-channel analogue of the barrier scattering matrix used in Ref.~\cite{borcea2024enhanced}.

\subsection{Right Cavity}

The right cavity is the mirror counterpart of the left cavity. 
We denote by $\mathcal{M}^\mathrm{R}$, $\mathcal{B}^\mathrm{R}$, $\mathcal{S}^\mathrm{R}$, and $\mathcal{C}^\mathrm{R}$ the port-index sets associated with the receive ports, barrier-facing ports, ``sensing'' ports, and configurational-diversity ports, respectively. 
The ordering within each set follows the mirror ordering used on the left. 
With this convention, the static mirror symmetry gives $\mathbf{S}^\mathrm{R}=\mathbf{S}^\mathrm{L}$; any difference between the reduced left and right cavity responses therefore originates only from the load vectors.

The sensing loads on the right define the unknown configuration, summarized by $\mathbf{r}^\mathrm{R}_\mathrm{S}$, whereas the sensing loads on the left define the programmable candidate configuration, summarized by $\mathbf{r}^\mathrm{L}_\mathrm{S}$. $\mathbf{r}^\mathrm{L}_\mathrm{S}$ can thus differ from $\mathbf{r}^\mathrm{R}_\mathrm{S}$.
By contrast, the configurational-diversity loads are always switched in mirror-symmetric pairs and are therefore described on both sides by the same vector $\mathbf{r}_\mathrm{C}$.
Analogous to \eqref{eq:left_reduced}, with $\mathcal{O}^\mathrm{R}=\mathcal{M}^\mathrm{R}\cup\mathcal{B}^\mathrm{R}$, $\mathcal{T}^\mathrm{R}=\mathcal{S}^\mathrm{R}\cup\mathcal{C}^\mathrm{R}$, and $\mathbf{\Phi}^\mathrm{R}=\mathrm{diag}([\mathbf{r}^\mathrm{R}_\mathrm{S},\ \mathbf{r}_\mathrm{C}])$, the reduced right-cavity scattering matrix is given by
\begin{equation}
\widetilde{\mathbf{S}}^{\mathrm{R}}
=
\mathbf{S}^{\mathrm{R}}_{\mathcal{O}^{\mathrm{R}}\mathcal{O}^{\mathrm{R}}}
+
\mathbf{S}^{\mathrm{R}}_{\mathcal{O}^{\mathrm{R}}\mathcal{T}^{\mathrm{R}}}
\mathbf{\Phi}^{\mathrm{R}}
\left(
\mathbf{I}_{N_\mathrm{T}}
-
\mathbf{S}^{\mathrm{R}}_{\mathcal{T}^{\mathrm{R}}\mathcal{T}^{\mathrm{R}}}
\mathbf{\Phi}^{\mathrm{R}}
\right)^{-1}
\mathbf{S}^{\mathrm{R}}_{\mathcal{T}^{\mathrm{R}}\mathcal{O}^{\mathrm{R}}}.
\label{eq:right_reduced}
\end{equation}

\subsection{Cascade of Subsystems}

As sketched in Fig.~\ref{Fig1}(c), the full system's $2M\times 2M$ scattering matrix $\mathbf{S}$ is obtained by cascading the reduced left cavity, the barrier, and the reduced right cavity via Redheffer star products (a definition of the Redheffer star product is provided in Appendix~\ref{Appendix_redheffer})~\cite{redheffer_inequalities_1959,anderson_cascade_1966,ha1981solid,prod2024updatable}:
\begin{equation}
\mathbf{S}
=
\widetilde{\mathbf{S}}^{\mathrm{L}}
\underset{\mathcal{B}^\mathrm{L},\mathcal{B}^{-}}{\star}
\mathbf{S}^{\mathrm{B}}
\underset{\mathcal{B}^{+},\mathcal{B}^\mathrm{R}}{\star}
\widetilde{\mathbf{S}}^{\mathrm{R}}.
\label{eq:total_cascade}
\end{equation}
The full system's transmission matrix from left to right can then be identified as
\begin{equation}
    \mathbf{T} = \mathbf{S}_{\mathcal{M}^\mathrm{R}\mathcal{M}^\mathrm{L}}.
\end{equation}

\section{Method}
\label{sec_Method}

The goal of our sensing technique is to determine the hidden configuration $\mathbf{r}^{\mathrm R}_\mathrm{S}$ of the $N_\mathrm{S}$ loads terminating the ``sensing'' ports of the right cavity. We can only measure the transmission matrix $\mathbf{T}$ from the left cavity's accessible ports in $\mathcal{M}^\mathrm{L}$ to the right cavity's accessible ports in $\mathcal{M}^\mathrm{R}$, while being able to tune the terminations of the left ``sensing'' ports in $\mathcal{S}^\mathrm{L}$ (summarized by $\mathbf{r}^{\mathrm L}_\mathrm{S}$) as well as the symmetric terminations of the left and right configurational-diversity ports in $\mathcal{C}^\mathrm{L}$ and $\mathcal{C}^\mathrm{R}$ (summarized by $\mathbf{r}_\mathrm{C}$).
The loads terminating the right ``sensing'' ports in $\mathcal{S}^\mathrm{R}$ can be open or short circuits. 
As in Ref.~\cite{yuan2026symmetry}, we deliberately impose the restriction to open and short circuits, and we deliberately keep the value of $N_\mathrm{S}$ modest. Thereby, we can exhaustively test all $2^{N_\mathrm{S}}$ possible termination configurations of the corresponding left ``sensing'' ports in $\mathcal{S}^\mathrm{L}$ and identify the one that is globally optimal for the optimization problem defined below in \eqref{eq:sensing_estimate}. Consequently, we do not need to worry about whether the optimization algorithm reaches the global optimum and can focus on the physics.

For a single choice of terminations of the $N_\mathrm{C}$ ports in $\mathcal{C}^\mathrm{L}$ and $\mathcal{C}^\mathrm{R}$, maximizing the total transmission at the operating frequency would not allow us to reliably identify $\mathbf{r}^\mathrm{R}_\mathrm{S}$: an asymmetric candidate configuration can happen to produce a narrowband resonance at, or very close to, the operating frequency and thereby yield a larger total transmission than the mirror-symmetric configuration.
This is precisely the issue that motivated the broadband averaging in Ref.~\cite{yuan2026symmetry}. 
Because each symmetric auxiliary-load configuration changes the cavity's interference landscape while preserving the target mirror symmetry, an accidental asymmetric resonance is unlikely to persist across many independent configurations, whereas the symmetry-induced enhancement is reinforced statistically across all configurations.
Here, we thus suppress such accidental resonant enhancements by averaging over configurational diversity instead of frequency diversity. 
For each candidate configuration $\mathbf{r}^{\mathrm L}_\mathrm{S}$, we therefore evaluate the total through-barrier transmission at a single frequency for $K$ different configurational-diversity states. 
In the $k$th realization, the auxiliary diversity loads (summarized by $\mathbf{r}^{(k)}_\mathrm{C}$) are chosen randomly but imposed symmetrically on the two sides of the barrier. 
Thus, the diversity loads change the scattering environment from one realization to the next without introducing additional mirror-symmetry defects, realizing symmetry-preserving configurational diversity. 

For each tuple $(\mathbf{r}^{\mathrm L}_\mathrm{S},\mathbf{r}^{\mathrm R}_\mathrm{S},\mathbf{r}^{(k)}_\mathrm{C})$, we compute the full system's transmission matrix $\mathbf{T}^{(k)}$ using the system model from Sec.~\ref{sec:system_model}. 
The corresponding total transmitted power is
\begin{equation}
\tau^{(k)}
\left(
\mathbf{r}^{\mathrm L}_\mathrm{S};
\mathbf{r}^{\mathrm R}_\mathrm{S}
\right)
=
\mathrm{Tr}
\left[
\mathbf{T}^{(k)}
\left(\mathbf{T}^{(k)}\right)^\dagger
\right].
\label{eq:tau_k}
\end{equation}
We then average this quantity over the $K$ diversity realizations:
\begin{equation}
\overline{\tau}
\left(
\mathbf{r}^{\mathrm L}_\mathrm{S};
\mathbf{r}^{\mathrm R}_\mathrm{S}
\right)
=
\frac{1}{K}
\sum_{k=1}^{K}
\tau^{(k)}
\left(
\mathbf{r}^{\mathrm L}_\mathrm{S};
\mathbf{r}^{\mathrm R}_\mathrm{S}
\right).
\label{eq:tau_average}
\end{equation}

Finally, we obtain our estimate of the hidden configuration as the mirror image of the left configuration $\mathbf{r}^\mathrm{L}_\mathrm{S}$ that maximizes the diversity-averaged total transmission. We identify
\begin{equation}
\widehat{\mathbf{r}}^{\mathrm L}_\mathrm{S}
=
\underset{\mathbf{r}^{\mathrm L}_\mathrm{S}}{\arg\max}
\;
\overline{\tau}
\left(
\mathbf{r}^{\mathrm L}_\mathrm{S};
\mathbf{r}^{\mathrm R}_\mathrm{S}
\right),
\label{eq:sensing_estimate}
\end{equation}
and obtain our estimate of the hidden right-side configuration as the mirror image of \(\widehat{\mathbf{r}}^{\mathrm L}_\mathrm{S}\); in the mirror-paired ordering introduced in Sec.~\ref{sec:system_model}, this means
\(\widehat{\mathbf{r}}^{\mathrm R}_\mathrm{S}=\widehat{\mathbf{r}}^{\mathrm L}_\mathrm{S}\).
To be clear, $\mathbf{r}^{\mathrm R}_\mathrm{S}$ is not assumed to be known to evaluate the objective function in \eqref{eq:sensing_estimate}; it only labels the fixed hidden configuration that determines the measured values of $\overline{\tau}$.

Our numerical implementation avoids recomputing the full cascade from scratch for every tuple. 
For each diversity realization, we first compute and cache the reduced scattering matrices of the cavity for all $2^{N_\mathrm{S}}$ candidate terminations of the ``sensing'' ports. 
These reductions are evaluated efficiently using Woodbury low-rank updates~\cite{prod2023efficient,prod2024updatable}. 
For each diversity realization, the baseline terminated-port problem is chosen with all ``sensing'' ports terminated in a reference state. 
A candidate sensing configuration that differs from this baseline in $h$ loads modifies the terminated-port linear system by a rank-$h$ update, so only an $h\times h$ auxiliary system has to be solved for that candidate.
During the scoring step, we precompute the blocks of the left-cavity--barrier cascade that enter the final transmission matrix, and then evaluate only the transmission block needed for $\mathrm{Tr}(\mathbf{T}\mathbf{T}^\dagger)$ rather than forming the full cascaded scattering matrix for every candidate pair. 
When we consider all possible right configurations in the reciprocal case, we further exploit the symmetry of the score matrix to evaluate only one triangular half of the candidate-pair scores. Specifically, the diversity-averaged score obtained for the candidate pair
$(\mathbf{r}^{\mathrm L}_\mathrm{S},\mathbf{r}^{\mathrm R}_\mathrm{S})$
is identical to that obtained after exchanging the two configurations,
$(\mathbf{r}^{\mathrm R}_\mathrm{S},\mathbf{r}^{\mathrm L}_\mathrm{S})$.
This symmetry follows from reciprocity together with the mirror-paired port ordering. Consequently, when all $2^{N_\mathrm{S}}$ right configurations are included, only one triangular half of the candidate-pair score matrix needs to be evaluated explicitly.
All of the above-described numerical implementation choices accelerate the exhaustive search without changing the objective in Eq.~(\ref{eq:tau_average}).

\section{Setup and Procedure}
\label{sec_SetupProcedure}

The D-shaped cavities shown in Fig.~\ref{Fig1}(a,b) have a height of $4~\mathrm{mm}$, a radius of $120~\mathrm{mm}$, and a cut-off segment height of $15~\mathrm{mm}$. Each cavity is equipped with $100$ SMA connectors that are randomly placed with a minimum distance of 8.5~mm. We obtained the corresponding $100\times 100$ scattering matrix at $6.27~\mathrm{GHz}$ from a numerical full-wave simulation in the 3D electromagnetic simulation software CST Studio Suite. Importantly, thanks to our MNT approach outlined in Sec.~\ref{sec:system_model}, only a single full-wave simulation is required for all results presented in this paper~\cite{tapie2024systematic}.
Unless stated otherwise, we use $M=4$, $N_\mathrm{B}=4$, $N_\mathrm{S}=8$, $N_\mathrm{C}\in\{14,24,34,44,54,64,74,84\}$, $t_\mathrm{B}=10^{-3}$, and $K\in\{1,2,4,8,16,32,64,128,256,512,1024,2048,4096,8192\}$. As described in Appendix~\ref{appendix_Anderson}, we have verified that our system is not in an effectively Anderson-localized regime at the operating frequency. 
In cases with $N<100$ (i.e., when $N_\mathrm{C}<84$), we work with a reduced version of the $100$-port scattering matrix obtained from CST: the unused ports are terminated with open-circuit loads and eliminated by applying the same MNT reduction procedure as in \eqref{eq:left_reduced} and \eqref{eq:right_reduced}. Accordingly, the matrices denoted by $\mathbf{S}^{\mathrm L}$ and $\mathbf{S}^{\mathrm R}$ in Sec.~\ref{sec:system_model} are the resulting reduced $N\times N$ cavity scattering matrices whenever there are unused ports.
We randomly partition the 100 SMA ports into the different types; we consider 10 random partitions and repeat our analysis for each of them.

\section{Results}
\label{sec_Results}

\begin{figure*}
    \centering
    \includegraphics[width=\textwidth]{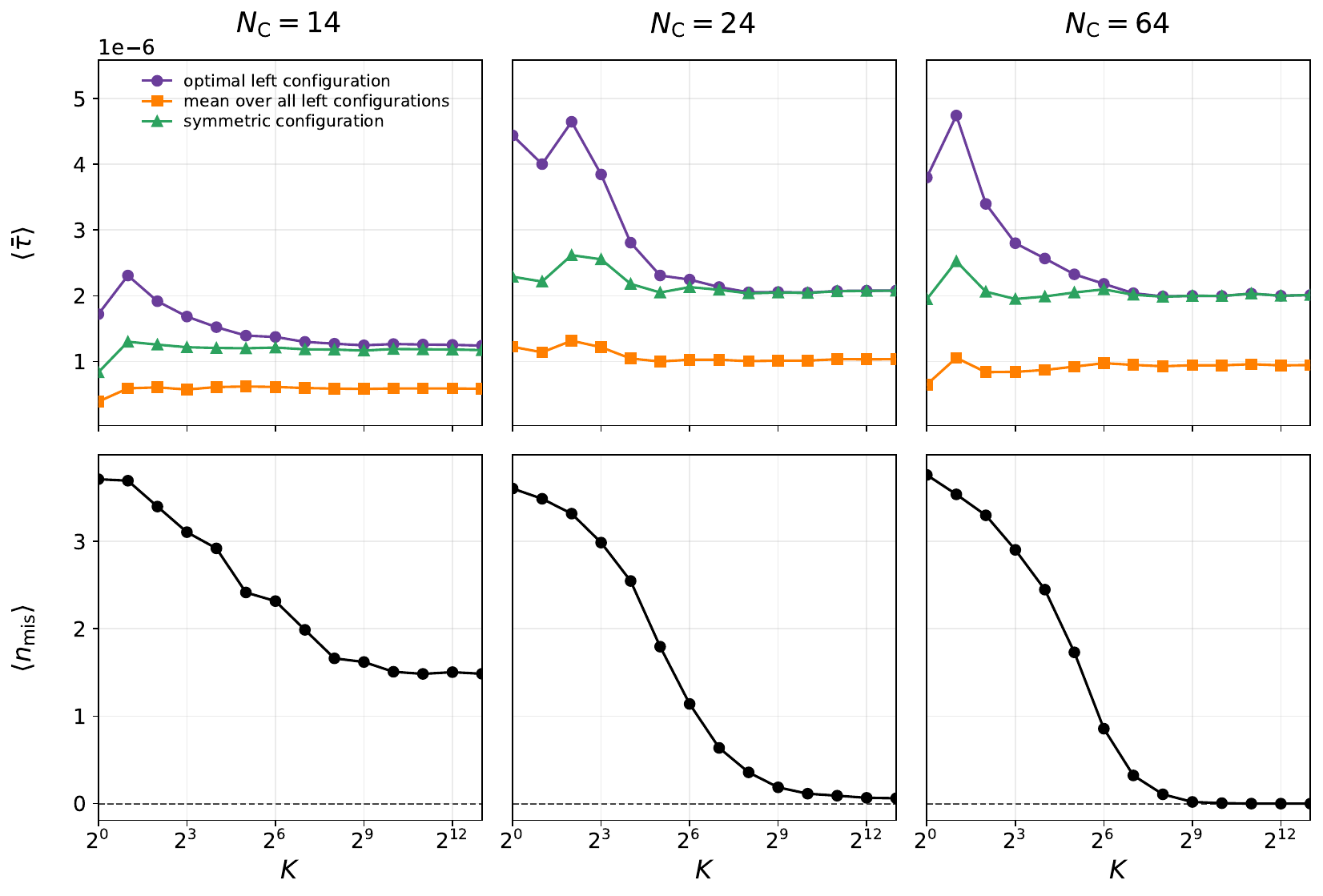}
    \caption{
Single-frequency sensing with configurational diversity. Columns correspond to $N_\mathrm{C}=14$, $24$, and $64$. Top row: ensemble average of the diversity-averaged transmission $\overline{\tau}$ as a function of $K$, comparing the optimal (purple), mean (orange), and mirror-symmetric (green) left sensing configurations. Bottom row: corresponding mean number of incorrectly sensed loads $\langle n_\mathrm{mis}\rangle$. Ensemble averages are over all 256 hidden right configurations and 10 random port partitions.
    }
    \label{Fig2}
\end{figure*}

We begin by examining the dependence of $\overline{\tau}$ on $K$ for different values of $N_\mathrm{C}$. In the top row of Fig.~\ref{Fig2}, we plot the mean of $\overline{\tau}$ over $2560$ realizations, corresponding to all possible $2^{N_\mathrm{S}}=256$ hidden right-side configurations to be sensed for each of the $10$ random port partitions. For each realization and each value of $K$, we compare three quantities: the value obtained by the globally optimal left sensing configuration (purple), the value obtained by the mirror-symmetric sensing configuration (green), and the mean value over all $2^{N_\mathrm{S}}$ possible left sensing configurations (orange). The configurational average over the $K$ symmetry-preserving auxiliary-load states is denoted by the overbar in $\overline{\tau}$, whereas the angular brackets in Fig.~\ref{Fig2} denote the additional ensemble average over the $2560$ sensing realizations. 
The green curves exceed the orange curves in all cases, indicating that the symmetric configuration systematically provides a notable enhancement over a random configuration. However, for small values of $K$, the green curves lie substantially below the purple curves, meaning that there exist asymmetric configurations that outperform the symmetric one. As $K$ increases, the gap between the purple and green curves shrinks. For large values of $K$, the gap between the green and purple curves becomes small for $N_\mathrm{C}=14$, it becomes visually imperceptible for $N_\mathrm{C}=24$, and it vanishes for $N_\mathrm{C}=64$. With sufficiently large $N_\mathrm{C}$, there is no asymmetric configuration that can outperform the symmetric one across a very large number of configurational realizations $K$. This observation is analogous to the one in Fig.~3 in Ref.~\cite{yuan2026symmetry}, where no asymmetric configuration outperformed the symmetric one across a very large number of independent frequency points.

To assess the sensing performance more directly, we next quantify the number of state variables for which the optimal left configuration differs from the mirror image of the unknown right configuration, i.e., the number of incorrectly identified loads. We denote this number by $n_\mathrm{mis}$ and plot its average over the 2560 realizations as a function of $K$ for the three considered values of $N_\mathrm{C}$ in the bottom row of Fig.~\ref{Fig2}. For all considered values of $N_\mathrm{C}$, $\langle n_\mathrm{mis} \rangle$ initially decreases with $K$ and then converges. For $N_\mathrm{C}=14$, $\langle n_\mathrm{mis} \rangle$ converges to a value of 1.48, meaning that the globally optimal configuration differs from the symmetric one by 1.48 entries on average. In the case of $N_\mathrm{C}=24$, $\langle n_\mathrm{mis} \rangle$ converges to 0.06. Although upon visual inspection the purple and green curves appear superposed for large $K$, the sensing performance is very good but not flawless: the globally optimal configuration differs from the symmetric one by 0.06 entries on average. This indicates that, in a few realizations, the symmetric configuration is not the globally optimal one for $N_\mathrm{C}=24$ with large $K$. In contrast, for $N_\mathrm{C}=64$, $\langle n_\mathrm{mis} \rangle$ converges to zero, indicating a flawless sensing performance.

Altogether, our results thus confirm the feasibility of single-frequency through-barrier sensing with sufficiently large $N_\mathrm{C}$ and $K$.
Moreover, since our system model is purely based on propagating waves, our results demonstrate that the symmetry-induced transmission enhancement and the associated sensing mechanism can arise without relying on evanescent coupling.

\section{Extension to Non-Reciprocal Cavities}
\label{sec_NonReciprocal}

In this section, we explore an extension to non-reciprocal cavities. Most previous works on symmetry-enhanced through-barrier transmission considered reciprocal systems, including the chaotic double-dot systems~\cite{macucci2007tunneling,whitney2009huge,totaro2010effect,macucci2020optimization}, graphene-based double dots~\cite{marconcini2013symmetry}, and diffusive disordered slabs~\cite{cheron2019broadband,cheron2020broadband,cheron2020sensitivity,davy2021experimental} mentioned in the Introduction. The symmetry-induced total-transmission enhancement originates from constructive interference between mirror-related multiple-scattering paths~\cite{whitney2009huge,cheron2019broadband,borcea2024enhanced,flegontov2025symmetry}. Hence, the essential requirement is not reciprocity itself, but rather that mirror-related paths contribute constructively through the mirror symmetry. This distinction is already visible in the quantum-dot setting of Ref.~\cite{whitney2009huge}, where a uniform magnetic field is considered as a perturbation that breaks the mirror-induced constructive interference and thereby suppresses the conductance peak (see Fig.~2(a) in Ref.~\cite{whitney2009huge}). Conversely, we hypothesize that if one could apply the negative of the left magnetic bias on the right side, the bias would be transformed together with the mirror operation, so that mirror-related paths would again accumulate equal phases.

We now consider an analogous idea in our microwave system. Instead of terminating all unused ports by open circuits, we terminate triplets of unused ports with ideal lossless circulators. Each circulator is connected to three ports of the same cavity and constitutes a non-reciprocal multiport termination. There are two possible circulator orientations. Assuming an ideal circulator without loss (which is approximately achievable at a single frequency), the corresponding $3 \times 3$ circulator scattering matrices are
\begin{equation}
\mathbf{S}_\circlearrowright =
\begin{pmatrix}
0 & 0 & 1\\
1 & 0 & 0\\
0 & 1 & 0
\end{pmatrix},
\qquad
\mathbf{S}_\circlearrowleft =
\mathbf{S}_\circlearrowright^\top
=
\begin{pmatrix}
0 & 1 & 0\\
0 & 0 & 1\\
1 & 0 & 0
\end{pmatrix}.
\label{eq:ideal_circulators}
\end{equation}
The first matrix routes waves as $1\rightarrow2$, $2\rightarrow3$, and $3\rightarrow1$, whereas the second matrix routes waves in the opposite direction. 
In the MNT reduction for the circulator-terminated cavities, we use the same elimination formula as in \eqref{eq:left_reduced} and \eqref{eq:right_reduced}, but generalize the termination matrices $\mathbf{\Phi}^\mathrm{L}$ and $\mathbf{\Phi}^\mathrm{R}$. Specifically, instead of being diagonal matrices of scalar load reflection coefficients, they become block-diagonal matrices whose blocks are the scalar reflection coefficients of the open-/short-circuit loads and the $3\times3$ scattering matrices of the ideal circulators.

\begin{figure*}
    \centering
    \includegraphics[width=0.9\textwidth]{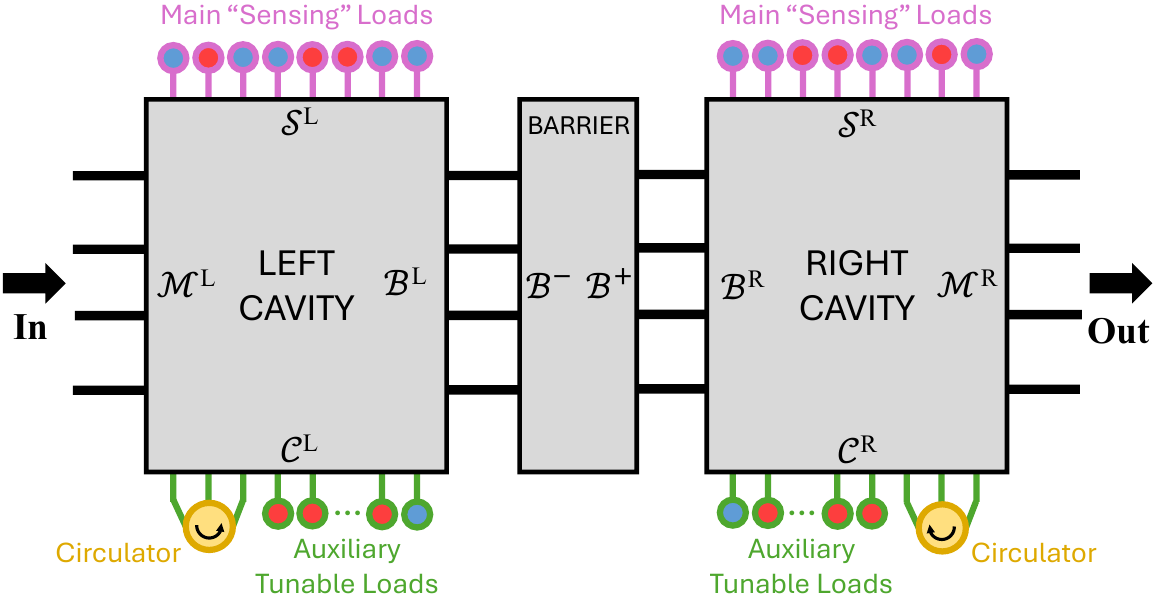}
	\caption{Modified MNT system model for the case in which unused ports are terminated by circulators rather than open circuits. The flipped orientation of the circulator originates from ``mirror-transformed circulator bias'' to preserve the global mirror symmetry.}
    \label{Fig3}
\end{figure*}

To preserve the constructive interference between mirror-related paths despite the presence of circulators, the relevant condition is that the effective scattering matrices of the two terminated cavities remain identical in the mirror-paired port ordering whenever all individual load states are identical, i.e., $\widetilde{\mathbf{S}}^{\mathrm L}=\widetilde{\mathbf{S}}^{\mathrm R}$. The physical meaning of this condition is sketched in Fig.~\ref{Fig3}. A circulator is a magnetically biased non-reciprocal element: its handedness is set by the sign of its internal magnetic bias. Under mirror reflection, this axial bias changes sign. Hence, the mirror image of a left-side circulator is not a circulator with the same laboratory-frame handedness, but one with the opposite handedness. We refer to this as ``mirror-transformed circulator bias''. As a benchmark, we also consider the non-mirror-transformed circulator-bias case, for which generally $\widetilde{\mathbf{S}}^{\mathrm L}\neq\widetilde{\mathbf{S}}^{\mathrm R}$ even when all individual load states are identical, because the generalized mirror symmetry is then deliberately broken.

\begin{figure*}
    \centering
    \includegraphics[width=0.5\textwidth]{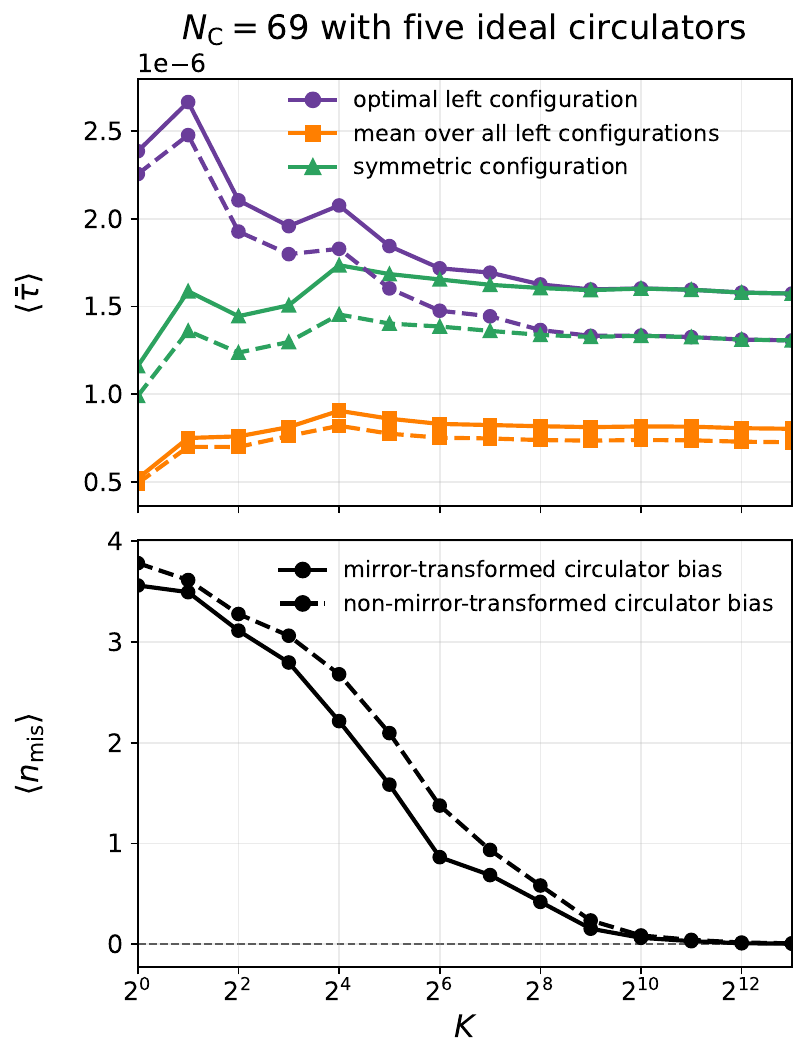}
    \caption{
Single-frequency sensing for $N_\mathrm{C}=69$ with ideal circulators terminating unused ports. Solid and dashed curves correspond to mirror-transformed and non-mirror-transformed circulator bias, respectively. Top: ensemble average of the diversity-averaged total transmission $\overline{\tau}$ as a function of $K$, comparing the optimal (purple), mean (orange), and mirror-symmetric (green) left sensing configurations. Bottom: corresponding mean number of incorrectly sensed loads $\langle n_\mathrm{mis}\rangle$. Ensemble averages are over all $2^{N_\mathrm{S}}=256$ hidden configurations and $10$ random port partitions.
    }
    \label{Fig4}
\end{figure*}

We consider a case with $N_\mathrm{C}=69$ in which we terminate the 15 unused ports of the left cavity with five circulators whose orientations are randomly picked. The mirror-symmetric ports of the right cavity are terminated with five circulators with ``mirror-transformed circulator bias'' or ``non-mirror-transformed circulator bias''. As before, we consider 2560 realizations (all possible 256 right hidden configurations, and ten random port partitions). When considering non-reciprocal systems in this section, we do not assume a symmetry of the score matrix to accelerate the exhaustive search. 
We quantify the strength of the non-reciprocity of each terminated cavity by evaluating $\|\widetilde{\mathbf{S}}^{q}-(\widetilde{\mathbf{S}}^{q})^{\top}\|_\mathrm{F}/
\|\widetilde{\mathbf{S}}^{q}\|_\mathrm{F}$, with $q\in\{\mathrm L,\mathrm R\}$. We find that this metric is around $0.24$--$0.27$ for each of the cavities. Thus, the five circulators introduce substantial non-reciprocity, which we expect since the long dwell time inside the cavity results in many scattering events involving the circulators such that the waves accumulate significant sensitivity to the non-reciprocal scattering within the circulators. This intuition is derived from analogous arguments for why tunable components impact a system's transfer function more when the waves have a longer dwell time~\cite{prod2025mutual,prod2025benefits}.

Analogous to Fig.~\ref{Fig2}, we plot in the top panel of Fig.~\ref{Fig4} the ensemble-averaged $\langle\overline{\tau}\rangle$ as a function of $K$ for both circulator-bias cases. The transmission is systematically larger for the mirror-transformed circulator bias, consistent with this case preserving the generalized mirror symmetry of the terminated cavities. For both cases, the purple and green curves converge as $K$ increases, indicating that the mirror-symmetric sensing configuration becomes globally optimal for (almost) all realizations at large $K$. Correspondingly, both curves of $\langle n_\mathrm{mis} \rangle$ approach zero for large $K$, with the mirror-transformed-bias case remaining slightly below the non-mirror-transformed-bias benchmark.
The non-mirror-transformed circulator bias behaves effectively like a symmetry-breaking ``noise'' source: it lowers the symmetry-enhanced transmission level, but its effect on the sensing decision is largely averaged out by the configurational diversity.

\section{Conclusion}
\label{sec:Conclusion}

To summarize, we have demonstrated a single-frequency variant of symmetry-empowered through-barrier sensing in which configurational diversity replaces the frequency diversity used previously in Ref.~\cite{yuan2026symmetry}. We tuned a set of main scatterers on the left side of the barrier to maximize the through-barrier total transmission averaged over a series of mirror-symmetric random configurations of auxiliary tunable scatterers on both sides of the barrier. For sufficiently large $N_\mathrm{C}$ and $K$, the globally optimal configuration of the left main scatterers becomes the mirror image of the sought-after unknown configuration of the right main scatterers. We also showed that the same framework naturally extends to non-reciprocal cavities: mirror-transformed circulator bias preserves the generalized mirror symmetry, whereas non-mirror-transformed bias reduces the enhanced transmission but is largely averaged out in the sensing decision.

Looking forward, the proposed single-frequency technique could be implemented either directly in an experiment, by optimizing programmable scatterers from measured transmission data, or indirectly in software, using an experimentally calibrated MNT model of the system. The latter route is particularly well aligned with our system model and could leverage recent progress in experimental MNT parameter estimation for programmable scattering systems~\cite{sol2024experimentally,del2025experimental,del2026ambiguity,del2026reduced,tapie2025experimental,tapie2026channel}.
For larger sensing spaces, arising either from a larger number of loads to be sensed or from multi-level or continuously tunable states of the loads to be sensed, exhaustive search across all possible configurations is no longer practical. Future work should therefore combine scalable optimization methods with criteria for assessing whether the retrieved configuration is symmetry restoring. Recent electromagnetic bounds on the achievable total transmission in programmable MIMO systems~\cite{del2026electromagnetic} may provide a useful benchmark for judging how closely an optimized configuration approaches the best physically achievable transmission.

\appendix

\section{Redheffer Star Product}
\label{Appendix_redheffer}

The forward Redheffer star product
${\mS}^{\U\V}=\mS^\U\underset{\mathcal{P}_\U,\mathcal{P}_\V}{\star}\mS^\V$
used in Eq.~(\ref{eq:total_cascade}) is defined by connecting the port set
$\mathcal{P}_\U$ of system $\U$ to the port set $\mathcal{P}_\V$ of system $\V$
and eliminating the corresponding internal wave amplitudes. 
The remaining, non-connected ports are denoted by $\iN_\U$ and $\iN_\V$.
The resulting scattering matrix ${\mS}^{\U\V}$ has blocks
\begin{align}
{\mS}^{\U\V}_{\iN_\U\iN_\U}
&=
\mS^\U_{\iN_\U\iN_\U}
-
\mS^\U_{\iN_\U\mathcal{P}_\U}
\mS^\V_{\mathcal{P}_\V\mathcal{P}_\V}
\mathbf{X}^{\U\V}
\mS^\U_{\mathcal{P}_\U\iN_\U},
\\
{\mS}^{\U\V}_{\iN_\U\iN_\V}
&=
-
\mS^\U_{\iN_\U\mathcal{P}_\U}
\mathbf{X}^{\V\U}
\mS^\V_{\mathcal{P}_\V\iN_\V},
\\
{\mS}^{\U\V}_{\iN_\V\iN_\U}
&=
-
\mS^\V_{\iN_\V\mathcal{P}_\V}
\mathbf{X}^{\U\V}
\mS^\U_{\mathcal{P}_\U\iN_\U},
\\
{\mS}^{\U\V}_{\iN_\V\iN_\V}
&=
\mS^\V_{\iN_\V\iN_\V}
-
\mS^\V_{\iN_\V\mathcal{P}_\V}
\mS^\U_{\mathcal{P}_\U\mathcal{P}_\U}
\mathbf{X}^{\V\U}
\mS^\V_{\mathcal{P}_\V\iN_\V},
\label{eq:redheffer}
\end{align}
where
\begin{align}
\mathbf{X}^{\U\V}
&=
\left(
\mS^\U_{\mathcal{P}_\U\mathcal{P}_\U}
\mS^\V_{\mathcal{P}_\V\mathcal{P}_\V}
-
\mI_{N_\star}
\right)^{-1},
\\
\mathbf{X}^{\V\U}
&=
\left(
\mS^\V_{\mathcal{P}_\V\mathcal{P}_\V}
\mS^\U_{\mathcal{P}_\U\mathcal{P}_\U}
-
\mI_{N_\star}
\right)^{-1},
\label{eq:redheffer_aux}
\end{align}
and $N_\star=|\mathcal{P}_\U|=|\mathcal{P}_\V|$ is the number of connected port pairs.

\section{Anderson Localization Sanity Check}
\label{appendix_Anderson}

In this Appendix, we summarize a sanity check to confirm that the selected operating frequency does not correspond to an effectively Anderson-localized regime. 
To this end, we examine the singular-value distribution of a single cavity's $4\times4$ transmission matrix $\mathbf{S}^\mathrm{L}_{\mathcal{B}^\mathrm{L}\mathcal{M}^\mathrm{L}}$, evaluated after terminating the remaining cavity ports by random binary open-/short-circuit loads. 
In the Anderson-localized regime, the rank of $\mathbf{S}^\mathrm{L}_{\mathcal{B}^\mathrm{L}\mathcal{M}^\mathrm{L}}$ would be approximately unity~\cite{shi2012transmission,davy2012focusing,wang2011transport,pena2014single,leseur2014probing}. 
For each load configuration, we compute the singular values $\sigma_i$ and the normalized transmission-eigenvalue weights $p_i=\sigma_i^2/\sum_j\sigma_j^2$. 
We then quantify the number of appreciably contributing transmission eigenchannels using the entropy-based effective rank $R_\mathrm{eff}=\exp(-\sum_i p_i\ln p_i)$~\cite{roy2007effective} and the eigenchannel participation number $P=1/\sum_i p_i^2$~\cite{davy2012focusing}. 
Both quantities equal unity for single-channel transmission and increase when several transmission eigenchannels contribute appreciably. 
We also count the number of significant singular values using the criterion $\sigma_i/\sigma_1>10^{-2}$. 
The statistics are obtained by pooling $100$ random binary load configurations for each of the $10$ independently chosen partitions of the 100 ports, corresponding to $1000$ single-cavity realizations in total. 
The resulting distributions confirm that the selected operating frequency is \textit{not} in an effectively Anderson-localized regime: the pooled diagnostic yields $3.9\pm0.2$ significant singular values, $R_\mathrm{eff}=2.36\pm0.36$, and $P=2.05\pm0.37$.

\begin{backmatter}

\bmsection{Funding}
This work was supported in part by the Nokia Foundation (project 20260028), the ANR France 2030 program (project ANR-22-PEFT-0005), the ANR PRCI program (project ANR-22-CE93-0010), and the Research Council of Finland (projects 371367 and 365679).

\bmsection{Disclosures}
The authors declare no conflicts of interest.

\bmsection{Data availability}
Data underlying the results presented in this paper may be obtained from the corresponding author upon reasonable request.

\end{backmatter}

%%%%%%%%%%%%%%%%%%%%%%% References %%%%%%%%%%%%%%%%%%%%%%%%%
%\bibliography{sample}

\providecommand{\noopsort}[1]{}\providecommand{\singleletter}[1]{#1}%

\end{document}